\documentclass[%
 reprint,
 amsmath,amssymb,
 aps,
]{revtex4-2}

\usepackage{graphicx}
\usepackage{dcolumn}
\usepackage{bm}
\usepackage{color}

\begin{document}

\preprint{APS/123-QED}

\title{Nonmagnetic-magnetic transition and magnetically ordered structure in SmS}

\author{S.~Yoshida}
\author{T.~Koyama}
 \altaffiliation[Present address:]{TAIYO NIPPON SANSO Corporation.}
\author{H.~Yamada}
\author{Y.~Nakai}
\author{K.~Ueda}
\author{T.~Mito}
 \email{mito@sci.u-hyogo.ac.jp}
 \affiliation{%
Department of Material Science, Graduate School of Science, University of Hyogo, Ako-gun 678-1297, Japan}
\author{K.~Kitagawa}
 \affiliation{%
Department of Physics, Graduate School of Science, University of Tokyo, Bunkyo-ku 113-0033, Japan}
\author{Y.~Haga}
\affiliation{%
Advanced Science Research Center,Japan Atomic Energy Agency,Tokai, Ibaraki 319-1195, Japan}

\date{\today}

\begin{abstract}
SmS,
a prototypical intermediate valence compound,
has been studied by performing high-pressure nuclear magnetic resonance measurements on a $^{33}$S-enriched sample.
The observation of an additional signal below 15-20~K above a nonmagnetic-magnetic transition pressure $P_{\rm c2} \approx 2$~GPa gives evidence of a magnetic transition.
The absence of a Curie-term in the Knight shift near $P_{\rm c2}$ indicates that
the localized character of $4f$ electrons is entirely screened and the mechanism of the magnetic ordering is not described within a simple localized model.
Simultaneously, the line shape in the magnetically ordered state is incompatible with a spin density wave order.
These suggest that the magnetic order in SmS may require an understanding beyond the conventional framework for heavy fermions.
The fact that hyperfine fields from the ordered moments cancel out at the S site leads us to a conclusion that the ordered phase has a type II antiferromagnetic structure.
\end{abstract}

\maketitle

Lanthanide-based semiconductors with a small temperature-dependent insulating gap, the so-called Kondo insulator including
SmB$_6$ and SmS, have continuously attracted the attention of researchers for half a century because of their rich and newly discovered fascinating properties.
In these materials, the relationship between the valence of lanthanide ions, magnetism, and transport properties has been a long-standing unsolved problem.
To investigate the issues, SmS may be an ideal material, since SmS exhibits several drastic phase transitions within a relatively narrow pressure range up to 2~GPa.
SmS, which crystalizes in the NaCl structure with almost divalent Sm ions, undergoes an isostructural valence transition at a pressure of $P_{\rm c1} = 0.65$~GPa \cite{Jayaraman}, above which an intermediate valence state evolves.
Further application of pressure causes a shift in the Sm valence toward the trivalent state, followed by a ground-state change from a nonmagnetic pseudogapped state to a magnetic metal at $P_{\rm c2} \approx 2$~GPa \cite{Barla,Haga,Matsubayashi}.
The sudden appearance of a hyperfine field (HF) at the Sm site \cite{Barla} and an anomaly seen in the thermal expansion \cite{Imura} at $P_{\rm c2}$ indicate the first-order character of this transition against pressure. 
The ground state above $P_{\rm c2}$ is expected to be antiferromagnetic from a decrease in the ac-magnetic susceptibility below an ordering temperature of $\sim 15$~K \cite{Matsubayashi_acX}.

One of the unique properties of SmS is that the Sm valence in the vicinity of $P_{\rm c2}$, estimated by high-energy x-ray experiments, is 2.6-2.8 \cite{Rohler,Deen,Annese,Imura_XAS}, far below the magnetic trivalent state.
Similar behavior is seen in SmB$_6$ \cite{Butch,Emi}, whereas it seems to differ from the cases in Ce and Yb systems where the nonmagnetic-magnetic transition occurs with nearly trivalent lanthanide ions \cite{Emi}.
A recent detailed study of the Sm valence in SmB$_6$ points out the duality lying in the valence deviation from the trivalent:
One is associated with low-energy valence fluctuations and the other with high-energy valence fluctuations \cite{Emi}.
Thus, the electronic states of the multi-$4f$ electron configuration ($4f^{5 - 6}$) in these Sm intermediate valence/Kondo insulators are largely unexplored.
More recently, SmS is
one of the few candidate materials
for a correlated topological insulator protected by time-reversal symmetry \cite{Li,Kang} as well as SmB$_6$ which has been proposed by earlier studies \cite{Dzero}.
In order to study various exotic phenomena arising from the magnetic effect on topological insulators, increasing interest has been
devoted to intrinsic magnetic topological insulators that exhibit stoichiometric magnetic ordering \cite{magTI}.
For SmS, the information on a magnetically ordered (MO) structure and the resultant microscopic distribution of the internal field, as well as the pressure dependence of the energy gap, will be crucial to clarify the relationship between the MO and the adjacent possible
topological states.

\begin{figure*}
\includegraphics[width=0.85\linewidth]{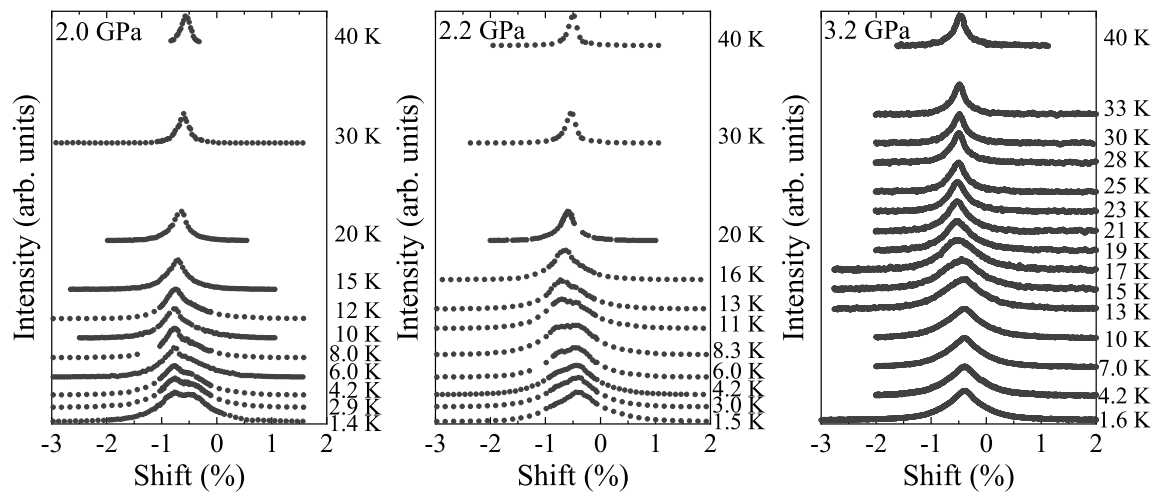}
\caption{\label{fig:wide}
$^{33}$S-NMR spectra at three different pressures plotted as a function of shift.
The shift
for the data
at 2.0 and 2.2~GPa, measured by sweeping frequency, was estimated as $2 \pi \nu_{\rm res} / \gamma_{\rm N} H_0 - 1$, while the shift at 3.2~GPa, measured by a sweeping field, was estimated as $2 \pi \nu_0/\gamma_{\rm N} H_{\rm res} -1$.
Here $\nu_{\rm res}$ and $H_{\rm res}$ are the resonance frequency and field, respectively, $H_0$ and $\nu_0$ are the applied constant field and frequency, respectively, and $\gamma_{\rm N}/2 \pi = 326.55$~Hz/Oe.
}
\end{figure*}

However,  there have been so far little
experimental data of
the neutron scattering and the nuclear magnetic resonance (NMR) 
on SmS, both of which are known as powerful techniques to determine the ordered structure, due to the facts that Sm is a neutron absorbing element and the natural abundance of the NMR-active isotope $^{33}$S is extremely low ($0.76 \%$).
To break through the situation, we prepared a $^{33}$S-enriched sample and made the first NMR report
on SmS at ambient pressure \cite{Koyama}.
The measurement of
NMR spectra provides static information on the HF,
and it can be performed under pressure.
Therefore, the Knight shift, estimated from the HF in a paramagnetic (PM) state and corresponding to the local susceptibility, is useful for investigating the pressure-induced evolution of magnetism.
The HF in an MO state directly connects with an ordered structure.
In this paper, the results of $^{33}$S-NMR measurements up to 3.2~GPa are shown.

A 98\% $^{33}$S-enriched powder sample of SmS was synthesized.
The preparation of the sample is described in Ref. \cite{Koyama}.
The $^{33}$S-NMR measurements were done at three different pressures above $P_{\rm c2}$.
The measurements at 2.0 and 2.2~GPa near $P_{\rm c2}$ were performed using a self-clamped BeCu/NiCrAl piston-cylinder cell with Daphne 7373 as a pressure medium.
The pressure was determined by a manganin gauge and a Sn manometer, both of which were placed inside the cell together with the sample and an NMR coil.
The application of a pressure of 3.2~GPa, much higher than $P_{\rm c2}$, was achieved using a modified opposed-anvil high pressure cell \cite{Kitagawa} with Daphne 7575 as a pressure medium.
The applied pressure was monitored using a Pb manometer and ruby fluorescence.
All the NMR experiments were carried out at the S site (the nuclear spin of $^{33}$S is 3/2) using a spin-echo technique with a phase-coherent pulsed spectrometer.
$^{33}$S-NMR spectra at 2.0 and 2.2~GPa were measured by sweeping frequency at a constant field of 6.0~T, while the spectra at 3.2~GPa were measured by sweeping field at a fixed frequency from 21.41 to 41.20~MHz.

Figure 1 shows the $^{33}$S-NMR spectra obtained at the three different pressures.
Here, the spectra are plotted against the shift, which allows us to compare the spectra measured at different magnetic fields.
Above 25~K at all pressures, regarded as in the PM phase, only a resonance line with a Lorentzian shape is observed.
Reflecting the high symmetry at the S site in SmS, the line shape is not influenced by the nuclear quadrupole interaction.
In contrast, as temperature decreases below 20~K, the $^{33}$S-NMR spectra especially for 2.0 and 2.2~GPa broaden and the line shape deviates from the single Lorentzian shape.

To see more about the changes in the line shape, some representative NMR spectra below 20 K  at 2.0 GPa are shown in Fig. 2(a).
The figure reveals that the single Lorentzian shape in the PM phase changes into a double-peak structure with decreasing temperature; more precisely, the PM signal slightly moves to a lower shift to reach $-0.8 \%$ at the lowest temperature, while an additional component emerges around $-0.4 \%$.
By considering the previously reported pressure-temperature phase diagram of SmS \cite{Barla,Haga,Imura}, the new component
is assigned to a signal from the MO phase stabilized by pressure.
Although the spectral changes at 3.2~GPa look less significant than around $P_{\rm c2}$,
the presence of magnetic order is certainly indicated by the detailed measurements of the shift and the spectral width
as described later and in the Supplemental Material (SM) \cite{Suppl}, and indeed the transition temperature is in good agreement with the results of a calorimetric experiment \cite{Haga}.
By decomposing the spectra using a double Lorentzian function, the volume fraction of each phase is obtained, because the
spectral intensity is proportional to the number of nuclei surrounded by the same local condition.
Fitting results are presented by the solid lines in Fig. 2(a).
In contrast, the spectra above 15 K at 2.0 GPa, above 17 K at 2.2 GPa, and above 21 K and below 13 K at 3.2 GPa are regarded to consist of a single component by examining the sum of squared residuals of the single Lorentzian fit (see the SM \cite{Suppl}).

\begin{figure}[t]
\includegraphics[width=0.9\linewidth]{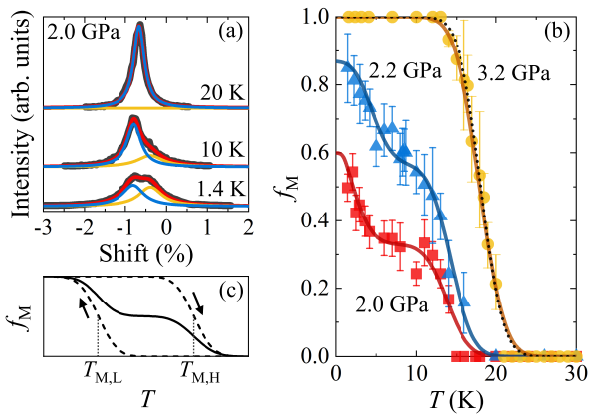}
\caption{\label{fig:epsart}
(a) Representative analyses of spectra measured at 2.0 GPa.
The red solid lines indicate least-squares fits of the data by the double Lorentzian function.
The blue and yellow lines represent the PM and MO components, respectively.
(b) Temperature dependence of $f_{\rm M}$ at 2.0, 2.2, and 3.2 GPa, which is obtained by the analysis demonstrated in (a).
The solid lines are the reproduction of the experimental data based on the thermal hysteresis model illustrated in (c) with Eqs.~(1) and (2).
The dotted line for the data at 3.2~GPa is the reproduction by assuming $T_{\rm M,H} = T_{\rm M,L}$.
The used parameters are listed in Table~I.
(c) Thermal hysteresis model of $f_{\rm M}$ with two different characteristic transition temperatures $T_{\rm M,H}$ and $T_{\rm M,L}$.
The dashed lines and the arrows indicate the heating and cooling processes.
The solid line is an $f_{\rm M}$-$T$ curve calculated by assuming that the two processes are mixed in a ratio of $1:1$.
See text for details.
}
\end{figure}

Figure~2(b) shows the temperature dependence of the volume fraction of the MO phase, $f_{\rm M}(T)$, at three different pressures.
One of the remarkable points is that the transitions at 2.0 and 2.2 GPa are broad, which is a manifestation of the coexistence of the PM and MO phases in wide temperature ranges.
The coexistence of the two phases near $P_{\rm c2}$ is consistent with the result of another microscopic measurement,  $^{149}$Sm-nuclear forward scattering (NFS) \cite{Barla}.
We should also note the unexpected temperature dependence of $f_{\rm M}$ at 2.0~GPa: After a rapid increase below 15 K, $f_{\rm M}$ shows a weaker temperature dependence below $\sim 10$~K, followed by another steep increase with cooling below $\sim 5$ K.
A similar tendency is also seen at 2.2 GPa, although $f_{\rm M}$ at the lowest temperature is much closer to 1 than at 2.0 GPa.

In order to understand the origin of the peculiar temperature dependence, one should first recall the previous suggestions that this transition is of first order \cite{Barla,Imura}.
The present fact that two distinguishable signals coexist in a wide temperature range is more evidence of the first-order character.
Then, we propose a model where $f_{\rm M}(T)$ follows a thermal hysteresis loop as illustrated in Fig. 2(c).
Here, the heating and cooling processes, i.e., $f_{{\rm M},i}$ vs $T$ curves having characteristic temperatures of transition $T_{{\rm M},i}$ ($i = $ H and L, respectively, and $T_{{\rm M,H}} > T_{{\rm M,L}}$), are described using a complementary error function:
\begin{eqnarray}
\label{eq1}
f_{{\rm M},i}(T) = \frac{1}{\sqrt{\pi}} \int^{\infty}_{T}e^{- t^2/2} dT,
\end{eqnarray}
where
$t = (T-T_{{\rm M},i})/\Delta T$.
In this model, we introduce a Gaussian distribution with a characteristic width of $\Delta T$ to the transition temperatures, which would otherwise be a step-function-like $f_{{\rm M},i}$-$T$ curve (see the SM for details \cite{Suppl}).
The point of this model is that the two states following the heating and cooling curves are assumed to coexist in the real powder sample, and therefore the total $f_{\rm M}$ is obtained as
\begin{eqnarray}
\label{eq2}
f_{\rm M}(T) = c f_{\rm M,H}(T) + (1-c) f_{\rm M,L}(T),
\end{eqnarray}
where $0 \leq c \leq 1$.
The transition between the two states may occur through an overlap due to the relatively large $\Delta T$ as well as temperature fluctuations before
starting the measurements \cite{T_fluctuation}.

\begin{table}[t]
\caption{\label{tab:table1}
Parameters in Eqs.~(1) and (2) used to reproduce the $f_{\rm M}$ data shown in Fig.~3(b).
For all the transitions, we used $\Delta T = 2.2$~K.
See text and Ref.~\cite{Suppl} for details.
}
\begin{ruledtabular}
\begin{tabular}{lllll}
$P$ & $T_{\rm M,H}$ & $T_{\rm M,L}$ & $c$ & $f_{\rm M}(0)$\\
\hline
2.0~GPa & 13.8~K & 1.5~K & 0.55 & 0.60 \\
2.2 & 14.3 & 4.2 & 0.65 & 0.87 \\
3.2 (solid line) & 19.0 & 16.9 & 0.50 & 1.0\\
3.2 (dashed line) & \multicolumn{2}{c}{18.1} & $\ -$ & 1.0 
\end{tabular}
\end{ruledtabular}
\end{table}

The solid lines in Fig.~2(b) show the reproduction by using the parameter values listed in Table~I.
Although this measurement does not track the hysteresis loop itself, the good agreement with the experimental results suggests that the unique temperature dependence of $f_{\rm M}$ is understandable within the scheme of the first-order transition.
At 2.0 GPa, $f_{\rm M}$ does not reach 1 at 0 K [$f_{\rm M}(0) = 0.6$].
This is also accounted for by considering the first-order character as a function of pressure;
namely, the PM and MO phases coexist at 0~K in the vicinity of $P_{\rm c2}$.
It is interesting to find that the marked thermal hysteresis at 2.0 GPa is largely suppressed with increasing pressure, as is evident from the pressure dependence of $|T_{\rm M,H} - T_{\rm M,L}|$ (see Table~I).
Here, the $f_{\rm M}$ data at 3.2 GPa is also well fitted by assuming $T_{\rm M,H} = T_{\rm M,L}$ as indicated by the dotted line in Fig.~2(b) and Table I.
Therefore it is possible that the hysteresis disappears at this pressure.
The suppression of the strong first-order character is consistent with the suggestion made by thermal expansion measurements that the phase boundary between the PM and MO phases changes from first order to second order with increasing pressure \cite{Imura}.

Next, the Knight shift $^{33}K$ estimated from the peak position of the spectra is demonstrated in Fig.~3(a).
For the temperature region where the PM and MO phases coexist, we used the results of the above-mentioned decomposition with the two Lorentzian functions.
In the PM phase ($T>20$ K), $|^{33}K|$ at all pressures monotonically increases with lowering temperature, however it is reduced with increasing pressure.
When the Sm valence shifts from the divalent state toward the trivalent one, the susceptibility is expected to be suppressed due to a decrease in predominant Van Vleck paramagnetic contributions.
This is the case for the intermediate valence state of SmS according to dc-susceptibility measurements up to around 1 GPa \cite{Matsubayashi_thesis,Imura_ThermalEx}.
The observed suppression of $|^{33}K|$ with pressure is also attributed to the same origin.

Note that $^{33}K$ at 2.0 GPa almost saturates below 10 K without showing any divergence.
The absence of the Curie term expected for the Sm trivalent component implies that the localized character of $4f$ electrons is screened through substantial hybridization near $P_{\rm c2}$.
This static property is consistent with the strong intermediate nature indicated by the x-ray absorption spectroscopy measurements \cite{Deen,Annese,Imura_XAS},
whereas it is distinctly different from the cases of Yb based heavy fermions in which
the evolution of the Curie term with pressure is seen near the nonmagnetic-magnetic transition \cite{YbXCu4}.
Hence the mechanism of the magnetic ordering is not described within a simple localized model.
Interestingly, the $|^{33}K|$ at 3.2 GPa increases more rapidly below 20 K and reaches or even surpasses the value at 2.0 GPa, which may be a sign of the evolution of the Curie term at higher pressures.

\begin{figure}[t]
\includegraphics[width=0.9\linewidth]{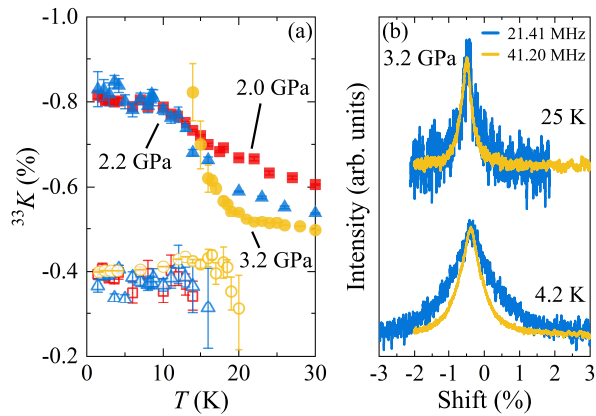}
\caption{\label{fig:epsart}
(a) Temperature dependence of $^{33}K$.
The solid and open symbols represent the estimation for the PM and MO states, respectively.
(b) Comparison of $^{33}$S-NMR signals in the PM (25 K) and the MO (4.2 K) phases measured at $\nu_0 = 41.2$ and 21.4 MHz.
The signals are plotted against the shift.
At 4.2 K, $H_{\rm res} \approx 12.66$ and 6.58 T for $\nu_0 = 41.2$ and 21.4 MHz, respectively.
}
\end{figure}

The resonance position in the MO phase is also plotted against $^{33}K$ in Fig.~3(a), revealing a considerable reduction in its absolute value compared to those for the PM signal.
Namely, the HF at the S nuclear position is reduced.
Here, according to the definition of the shift (see the caption of Fig.~1), if the HF is caused by spontaneous magnetization which is independent of external field, the estimated value will depend on $H_0$ or $\nu_0$.
However, this is not the case for the present result as shown in the lower panel of Fig.~3(b):
The shifts at the peak of the spectra measured at different NMR frequencies coincide.
Moreover, note that a powder pattern spectrum in an MO phase generally has a trapezoidal shape due to the random distribution of the local internal field with respect to the external field.
In contrast, the observed spectra in the MO phase retain the Lorentzian shape [see the lower panel of Fig.~3(b)], distinguishable from the trapezoidal one.
These results consistently give evidence that the HF's originating from the MO moments cancel out at the S site.
As the MO state is metallic \cite{Holtzberg,Lapierre}, the residual $^{33}K$ ($\sim -0.4$ \%) may be ascribed to the contribution of conduction electrons.

Here, we comment on the spectral width in the MO phase at 3.2 GPa.
It is 85-90 mT and independent of $\nu_0$ as shown in Fig.~S3(a) of the SM \cite{Suppl},
indicating that
the width is dominated by spontaneous
HF.
This also results in the ``seeming'' broadening when plotting spectra measured at lower fields against the shift, as shown in the lower panel of Fig.~3(b).
Such a phenomenon is not detected in the PM state [see the upper panel of Fig.~3(b)].
The spectral width in the MO phase is comparable with those of the extracted PM spectral component around $T_{\rm M,H}$ ($\sim 80$ mT) \cite{Suppl}.
If the MO structure is spin-density-wave (SDW)-like and the resulting oscillatory HF is the main cause of the linewidth, the spectra would be much broader than those in the PM phase.
Thus, the occurrence of commensurate/incommensurate SDW order seen in itinerant magnetic systems
is unlikely.
Alternatively, the field-independent line width is simply attributable to the HF distributed around zero, which may arise from the sample inhomogeneity.

\begin{figure}[t]
\includegraphics[width=0.9\linewidth]{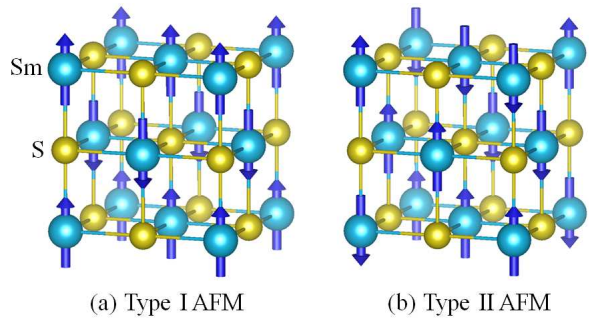}
\caption{\label{fig:epsart}
(a) Type I AFM and (b) type II AFM structures, drawn using VESTA \cite{VESTA}.
The latter is the plausible structure for the MO phase of SmS.
It is not possible to determine the direction of the ordered moments in this experiment.
}
\end{figure}

We have investigated a plausible MO structure that satisfies the present experimental results.
Of the simple type I and II antiferromagnetic (AFM) structures generally found in rare-earth monochalcogenides (see Fig.~4) \cite{Vogt},
in the case of a type II structure, all pairs of Sm atoms at symmetric positions with respect to a given S site have an oppositely polarized moment, resulting in the cancellation of the internal field at the S site.
In contrast, a nonzero internal field at the S site is evident in the case of a type I structure by considering the superposition of dipole fields from the surrounding Sm moments.
Therefore the type II AFM structure is most likely to be realized in the MO phase.
Note that the HF at the Sm site is as large as 300~T, probed by a $^{149}$Sm-NFS experiment \cite{Barla}, whereas the present study indicates that it vanishes at the S site due to high symmetry in the ordered structure.
The microscopic information will be useful for further investigations of electronic states in the MO phases.

In summary, we have investigated the nonmagnetic-magnetic transition in SmS by means of $^{33}$S-NMR measurements using a $^{33}$S-enriched sample.
The occurrence of magnetic ordering is evidenced from the observation of two distinguishable signals
at 2.0 and 2.2 GPa near $P_{\rm c2}$, where
the localized character of $4f$ electrons in the PM component
is entirely screened probably through the substantial hybridization,
which is
in sharp contrast to the behavior in Yb-based heavy fermions.
Simultaneously, the line shape in the MO state is incompatible with an SDW order.
These suggest
that the magnetic ordering in SmS may require an understanding beyond the conventional framework for heavy fermions.
The temperature and pressure dependences of $f_{\rm M}$ are well described using the thermal hysteresis model, whereas the strong first-order character is suppressed at 3.2 GPa.
The present study also reveals that, in the MO phase, the HF's at the S site cancel out, which leads us to a conclusion that the MO state has a type II AFM structure.

\begin{acknowledgments}
We are grateful to Prof. Y.~Hasegawa and Prof. K.~Matsubayashi for valuable discussions.
This work was supported by JSPS KAKENHI (Grants No. 16K05457, No. 18H04331, and No. 15H05883).
\end{acknowledgments}


\begin{thebibliography}{99}
\bibitem{T_fluctuation}
The temperature fluctuations during the measurements are small, i.e. less than $\pm0.1$~K.
Therefore the fluctuations until starting the measurements may be responsible for causing the transitions between the two phases through the overlap due to the relatively large $\Delta T$.
\end{thebibliography}

\end{document}